\documentstyle[aps,prb,epsf,preprint]{revtex}
\begin{document}



\title{Quantum Phase Transitions and Correlated Electrons}

\author{Qimiao Si}
\address{Department of Physics \& Astronomy, Rice University, Houston,
TX 77005--1892, U.S.A. }

\maketitle

\begin{abstract}

\vskip 0.5 in

{\bf 

This article is aimed at a pedagogical introduction to
the physics of quantum phase transitions that is unique
to metallic systems.
It has been recognized for some time that quantum criticality 
can result in a breakdown of Landau's Fermi liquid theory.
Its converse, however, has not been appreciated until
very recently: non-Fermi liquid behavior can in turn
lead to new classes of quantum phase transition.
A concrete example is provided by ``local quantum critical points''.
I summarize the theoretical reasoning and experimental evidence
for local quantum criticality, in the context of heavy fermion
metals. The underlying physics is likely to be
relevant to other correlated electron systems including 
doped Mott insulators.
}

\end{abstract}

\newpage

It would seem hopeless to describe the physics of $10^{22}$
or so electrons which are strongly interacting with each other.
Yet, for a long time, a remarkably simple theory was considered
to be the definite solution.
First formulated almost half a century ago,
Landau's Fermi liquid theory was successfully applied 
not only to simple metals but also to a number of systems
in which interaction effects are very strong. For instance,
it appears to work even for systems with an effective 
electron mass that is a few hundreds of the corresponding
band-structure value. Over the past decade or so, however, 
a long list of materials
have emerged in which the Fermi liquid description apparently
fails. In addition to high temperature superconductors,
examples in this category include the $f$-electron-based heavy
fermion metals, $d$-electron-based transition-metal compounds
(beyond the cuprate oxides), and quasi-one-dimensional materials 
such as single-walled carbon nanotubes and semiconductor 
quantum wires. The basic question we are confronted with
is when and how electron-electron interactions lead to a breakdown
of the Fermi liquid theory. 

A number of mechanisms for non-Fermi liquid behavior are currently
being pursued. Here we are concerned with one of these, namely
proximity to a quantum critical point. In the remainder of this
pedagogy-minded article, I will i) briefly introduce 
quantum phase transitions in correlated electrons and its links
to non-Fermi liquid phenomena, 
ii) summarize some recent progresses in this area,
with an emphasis on the notion of local quantum critical points
and iii) comment on some broader implications of these results.
For further readings, I refer the readers to the following
books and articles:
Ref.~\onlinecite{Sachdev-book} for a recent 
review on quantum phase transitions in general; 
Refs.~\onlinecite{Varma,Si-Nature,JPCM,Hertz,Millis}
for quantum critical points in correlated electrons;
Refs.~\onlinecite{Shankar,Nozieres} for 
Fermi liquid theory;
Refs.~\onlinecite{Hewson,Anderson61} for local moment
formation and Kondo effects;  
Refs.~\onlinecite{Stewart,Schroeder,Steglich,Steglich2,Aronson,Lohneysen,Mathur}
for experiments
in quantum critical heavy fermions and
Refs.~\onlinecite{Varma,Johnson}, which serve as points of departure
towards a large literature arguing for 
quantum criticality
in high temperature cuprate superconductors.
A recent review (ref.~\onlinecite{Si0211})
discusses similar issues at a more technical level and 
contains more references to original works.

\section{Quantum criticality meets non-Fermi liquids}

\subsection{Fermi liquid theory and its breakdown}

The basic ingredient of the Fermi liquid theory is 
that the elementary excitations of
an interacting many-electron system are quasiparticles.
These quasiparticles are 
sufficiently long-lived at low energies, and have
the same intrinsic quantum numbers as a bare electron
(spin ${\hbar \over 2}$ and charge $\pm e$, with $-e$ for
quasiholes -- see below).

\begin{figure*}[h]
\centerline{\epsfysize=10cm\epsfbox{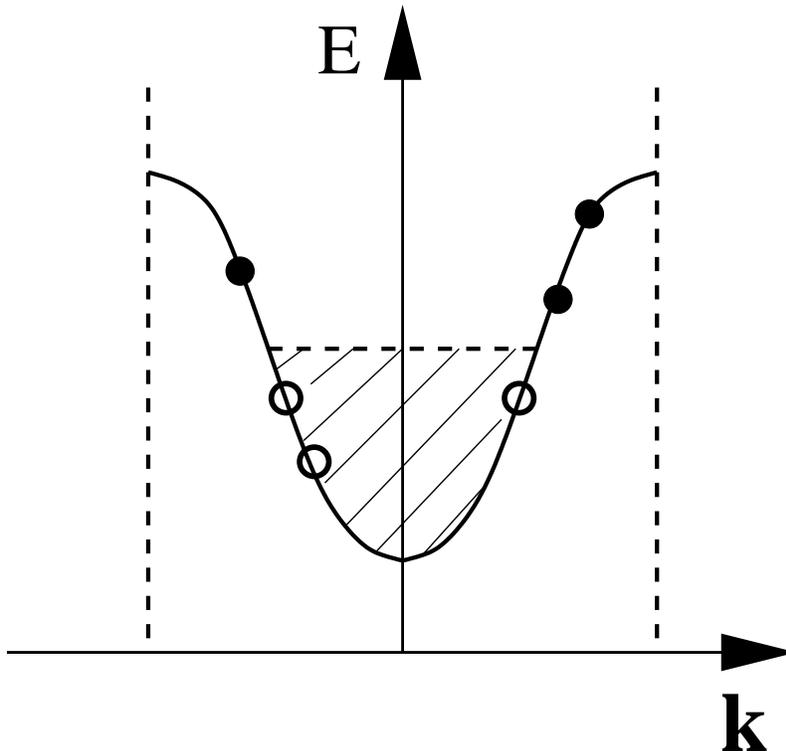}}
\caption{Elementary excitations of a non-interacting electron system.
The horizontal dashed line marks the Fermi energy and the vertical
dashed lines specify the boundaries of the first Brillouin zone along
a particular direction in the wavevector space. 
For illustrative purposes, only one energy band in the vicinity of
the Fermi energy is shown. 
}
\label{non-interacting}
\end{figure*}

The meaning of an elementary excitation can be best illustrated 
by considering a system of $N$ (of the order of $10^{22}$)
non-interacting electrons in a periodic potential.
The quantum mechanical eigenstates
of such an ideal electron system can be specified in terms
of those for a single electron in the same periodic potential.
The latter, as explained in any elementary solid state textbook,
are Bloch states that are characterized by 
a ``crystal wavevector'' ${\bf k}$ and a band
index. These parameters specify the bandstructure
as illustrated in Fig.~\ref{non-interacting}.
The intrinsic quantum numbers of these Bloch states
remain charge $e$ and spin $\pm {\hbar \over 2}$.
The eigenstates of the $N$-ideal-electron system are 
simply Slater determinants of 
the Bloch states.
The many-body ground state is the Slater determinant of
$N$ lowest-energy Bloch states; this is the filled Fermi sea
shown in Fig.~\ref{non-interacting}.
The many-body excited states
can be constructed by moving $m$ ($<N$) electrons from 
Bloch states in the Fermi sea to those above 
the Fermi energy.
Pictorially, an excited state wavefunction can be constructed
by digging $m$ ``holes'' (an empty circle in 
Fig.~\ref{non-interacting}
below the Fermi energy) 
and adding $m$ ``particles'' (filled dots in 
Fig.~\ref{non-interacting} above the Fermi energy).
Each ``particle'' or ``hole'' is then an elementary excitation.
It is obvious from this construction that our elementary
excitations 
have 
charge $\pm e$ and spin $\pm {\hbar \over 2}$.

Real electrons in solids of course do interact with each other.
Once these interactions are included, the many-body spectrum
becomes hard to construct exactly (except for simplified models
in one spatial dimension). 
The assumption of the Fermi liquid theory is that, the 
{\it low-lying} many-body excited states can still be constructed
from the many-body ground state by simply
adding
quasiparticles and quasiholes.
Compared to the dots and holes of Fig.~\ref{non-interacting}, 
a quasiparticle is much more complex. It can be pictured as
a bare electron with a polarization cloud attached.
Nonetheless, in many ways a quasiparticle/quasihole behaves just
like a bare electron/hole in a non-interacting electron system:
it has charge $\pm e$ and spin $\pm {\hbar \over 2}$, and it 
obeys Fermi statistics.
The lower the energy,
the more precise this construction is, as reflected in
the quasiparticle lifetime that goes to infinity as the wavevector
approaches the Fermi surface. 

These phenomenological statements can be proven by treating the 
electron-electron interactions perturbatively, to infinite orders.
(Unless special bandstructure effects such as nesting come into 
play, the only perturbative instability occurs when the effective
interaction in some pairing channel is attractive; the resulting 
state is the celebrated, and well-understood, BCS superconductor.)
In this sense, 
the Fermi liquid theory is internally consistent
when the
interactions 
are not too strong.

But how strong is too strong? The answer to this question is 
{\it a priori} unknown.

\subsection{Quantum phase transitions}

Just like ice melts as the thermal fluctuations are increased with
increasing temperature, an ordered state (such as an antiferromagnet)
at zero temperature can become disordered 
as the zero-point motion is tuned. 
The latter can be achieved through varying an external parameter 
such as pressure. The vanishing of
the order parameter characterizes a quantum
phase transition.
For definiteness and also 
with the heavy fermion metals
in mind, we will take the ordered state 
to be an antiferromagnetic metal; the corresponding order parameter
is the staggered magnetization. With appropriate modifications, 
the discussions below would apply to other types of ordered states.
There are also quantum phase transitions
for which
neither of the phases has an obvious order parameter,
as exemplified by the metal-insulator transitions 
in disordered interacting electron systems;
we will not consider such situations here.

A quantum critical point arises when the zero-temperature transition
is continuous. Here, the order parameter fluctuations are critical.
In the magnetic case, the spin susceptibility diverges in the limit of
zero frequency and zero temperature and as the wavevector approaches
the ordering wavevector.

\subsection{The connection between a quantum critical point and 
non-Fermi liquid}

Quantum critical points are especially complex in correlated
electron systems.
Besides the order parameter fluctuations, we also have to deal with
the electronic excitations near the Fermi energy.
A key observation, known 
since the original work of John Hertz,
is that the effective interactions between the electrons 
can be infinite due to the divergence of the order parameter
susceptibility. 
Such a divergence invalidates the aforementioned condition 
of internal consistency for the Fermi liquid theory,
opening the door to a non-Fermi liquid critical state. 

The effective electron-electron interactions become finite away from
the quantum critical point.
In principle, Fermi liquid theory may break down before the onset
of magnetic order. For the systems under discussion,
it is generally believed theoretically and supported experimentally
that the system remains a Fermi liquid on the paramagnetic side
and turns into a non-Fermi liquid only at the quantum critical point.
In other words, the emergence of non-Fermi liquid behavior 
coincides with the onset of magnetic ordering.

Such a linkage between quantum critical points and non-Fermi liquids 
has two important sides to it.
On the one hand, it already says that 
quantum criticality
provides a mechanism for non-Fermi liquid behavior.
Experimentally, the situation is especially clear-cut in heavy
fermion metals.
In the quantum critical regime, the electrical resistivity 
is linear (or close to being linear) in temperature;
the specific heat divided by temperature either diverges as temperature
goes to zero or at least has a non-analytic dependence on temperature.
Away from the quantum critical regime,
both the electrical resistivity 
and specific heat coefficient recover the Fermi liquid form
($T^2$ and constant, respectively).

On the other hand, the fact that the Fermi liquid - non-Fermi liquid
transition coincides with the magnetic phase transition raises
a more drastic possibility: the non-Fermi liquid physics can in
turn change
the universality class of the quantum phase transition itself.
This converse effect has not been discussed until recently.

\section{Gaussian quantum critical metals}

To appreciate how non-Fermi liquid behavior  can modify the 
quantum critical
physics, we first outline the 
standard theory of metallic quantum critical points.
This picture 
was introduced in the modern form by
John Hertz in 1975. It can be traced back even further,
to the literature on paramagnons in the 1960's;
the field theoretical formulation most commonly used today is
referred to as the Hertz-Millis theory.
In spite of its non-Fermi liquid nature, the electronic states
are not considered as a part of the critical theory;
instead, they are treated as bystanders.
The only critical modes are 
the long-wavelength
fluctuations of the magnetic order parameter - the paramagnons.
The critical theory describes the non-linear couplings of
such paramagnons, and assumes the form of 
the standard ``$\phi^4$'' theory with an effective dimensionality 
of $d_{eff} = d + z$. The effective dimensionality is raised from
the spatial dimension $d$ by $z$, the dynamic exponent, reflecting
the mixing of statics and quantum dynamics.
The primary effect of the bystanding electrons is to provide
extra channels for the critical spin fluctuations to decay into,
leading to an over-damped situation: the spin-damping term
is the strongest term in the frequency dependence.
This makes the dynamic exponent
$z$ larger than one -- $z=2$ for the antiferromagnetic case.
As a result, in three or two spatial dimensions
$d_{eff}$ is larger than or equal to $4$, the upper
critical dimension of the $\phi^4$ theory. The critical theory
is then Gaussian, and 
physical properties are expected to have 
a very simple 
mean-field behavior. 
For instance, the dynamical spin
susceptibility is expected to be linearly dependent on the 
frequency $\omega$.

Can the non-Fermi liquid electronic excitations really be taken
as bystanders? And would such electronic excitations directly
participate in the critical theory leading to a breakdown of the
Gaussian picture?
These turned out to be not just
academic questions. 
The past few years has witnessed a systematic experimental test of 
the Gaussian picture. 
The experiments have become possible because quantum critical
points have been explicitly identified in a number of heavy fermion
metals,
including ${\rm Yb Rh_2Si_2}$, ${\rm CeCu_{6-y}Au_y}$,
${\rm CePd_2Si_2}$ and ${\rm CeIn_3}$.
Inelastic neutron scattering experiments, particularly those
of Almut Schr\"oder and co-workers,
have raised the most striking puzzles:

\begin{itemize}

\item The frequency and temperature dependences of the dynamical spin
susceptibility 
display an anomalous exponent, $\alpha < 1$,
as well as $\omega/T$ scaling. 
The critical theory
cannot be Gaussian.

\item
The same anomalous exponent $\alpha$ is seen essentially everywhere
in the 
wavevector space,
suggesting new local physics.

\end{itemize}

The non-Gaussian 
and local nature 
calls for new 
critical physics beyond the paramagnons.

\section{Local quantum critical metals}

To address the interplay between the onset of magnetic ordering and 
the development of non-Fermi liquid behavior,
we have to go into some microscopics about the electronic excitations.
Our focus is on
the heavy fermion metals, though the discussion may be extended to
other strongly correlated electron systems as well.
The name ``heavy fermions'' itself refers to the fact that materials
containing partially-filled $f$-orbitals often behave as though 
the electrons have a very heavy mass (typically a few hundred times
of the value given by bandstructure calculations).

\begin{figure*}[h]
\centerline{\epsfysize=10cm\epsfbox{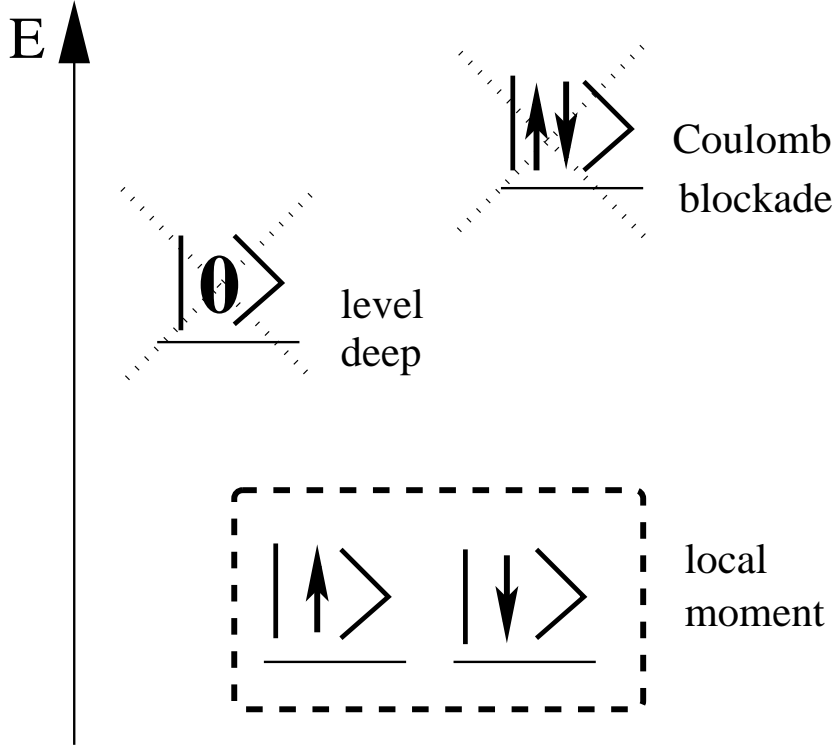}}
\caption{Local moment formation.
}
\label{local-moment-formation}
\end{figure*}

On the paramagnetic side, the formation of a heavy Fermi liquid
-- and the associated heavy quasiparticles -- is an intricate many-body
process. A microscopic Hamiltonian, appropriate for heavy fermions
at low energies (typically 100 K and below) is the Kondo lattice model:
a lattice of spin-${\hbar \over 2}$ local magnetic moments,
which are only weakly coupled to a separate conduction electron band. 
(There are seven $f$-orbitals, but at the energy scale of interest
we can focus on only one of them -- the lowest Kramers doublet.)
The formation of local magnetic moments itself is the result of very 
strong Coulomb interactions between two electrons occupying the same
$f$-orbital. Double occupancy of an $f$-orbital is excluded as a result
of a microscopic Coulomb blockade. The empty orbital is also
unfavorable when the $f$-orbital is a deep level ($i.e.$ its energy
is much lower than the chemical potential). Only spins 
are left as the low-lying degrees of freedom,
as illustrated in Fig.~\ref{local-moment-formation}.

\subsection{Kondo resonance}

The electronic excitations of a heavy Fermi liquid are composed
of Kondo resonances. 
The notion of Kondo resonance is most clearly understood
in the case of a single-impurity Kondo problem.
The Kondo resonance appears in the $f$-electron spectral function
as a peak in the vicinity of the Fermi energy. The width
of this peak is of the order of $kT_K^0 \sim k \rho_0^{-1}
{\rm e}^{-1/\rho_0 J}$, where 
$\rho_0$ is the conduction electron density of states at the
Fermi energy and $J$ the antiferromagnetic Kondo exchange coupling
between the magnetic moment and spin of the conduction electrons.
The center of the peak is in the vicinity of the Fermi energy,
with a separation that is smaller than $kT_K^0$.

The key to the existence of a Kondo resonance is the
singlet nature of the ground state. This can already be seen
in a simple atomic Kondo problem. Consider a spin-${\hbar \over 2}$ 
local moment (call it ${\bf S}$) coupled to a single conduction
electron orbital (call it $a$), as specified by the Hamiltonian
\begin{eqnarray}
H = J {\bf S} \cdot \sum_{\sigma,\sigma'} 
a_{\sigma}^{\dagger} 
{ { \vec{\bf \tau}_{\sigma,\sigma'} } \over 2} a_{\sigma '}
+ \epsilon_a \sum_{\sigma} a_{\sigma}^{\dagger} a_{\sigma} ,
\label{1+1}
\end{eqnarray}
where $\vec{\bf \tau}$ represents the Pauli matrices.
Such a Kondo Hamiltonian arises by projecting 
the Hilbert space of the impurity $f-$orbital to 
the low-energy local-moment subspace as already shown
in Fig.~\ref{local-moment-formation}.
This projected atomic problem has a reduced Hilbert
space, of eight dimensions.
For our illustrative purpose, it suffices to consider 
the case $\epsilon_a=0$.
(The Kondo lattice model we will be considering is
particle-hole asymmetric; but this distinction is 
unimportant at the moment.)
The ground state is a singlet:
\begin{eqnarray}
|0> = {1 \over \sqrt{2}} (|\uparrow>_f |\downarrow>_a
-
|\downarrow>_f |\uparrow>_a) 
\label{singlet}
\end{eqnarray}
The ket with a subscript $f$ ($a$)
acts in the local-moment (conduction electron) space.
The first low-lying exited states are two doublets,
\begin{eqnarray}
|\sigma>_1 && = |\sigma>_f |0>_a \nonumber\\
|\sigma>_2 && = |\sigma>_f |2>_a
\label{doublets}
\end{eqnarray}
To determine the $f$-electron spectral function, 
we need to know the form of the $f$-electron
operator in the projected local-moment subspace
(again, c.f. Fig.~\ref{local-moment-formation}).
A relatively straightforward calculation (through
the same Shrieffer-Wolff canonical transformation that
leads to the Kondo Hamiltonian Eq.~\ref{1+1}) shows 
that this projected operator reads
\begin{eqnarray}
F_{\sigma} = 
-\sqrt{{2J\over {U}}}
[\sigma S_z a_{\sigma} +
(S_x - \sigma i S_y) a_{-\sigma}]
\label{F-sigma}
\end{eqnarray}
The parameter $U$ is the original on-site Coulomb interaction
between two $f$-electrons; this parameter also characterizes the spacing
between the singly-occupied and doubly occupied levels shown
in Fig.~\ref{local-moment-formation}.
It is easily seen that 
$F_{\sigma} |0> = \sigma {3 \over 2}\sqrt{{J \over U}}
|-\sigma>_1$ and $F_{\sigma}^{\dagger} |0> = 
{3 \over 2}\sqrt{{J \over U}} |\sigma>_2$.
This finite matrix element of the $f$-electron creation/annihilation 
operator between the singlet ground state and the first excited
doublets establishes the existence of a small amount of the 
$f$-electron spectral weight at low energy, of order $J$ as opposed to $U$. 
This excitation is precisely the 
Kondo resonance as manifested in this atomic model; when we go beyond
the atomic limit, the scale $J$ turns into the Kondo scale $T_K^0$.
Such low-lying excitations would be absent if the ground state were not 
a singlet. 


\subsection{Destruction of the Kondo resonance}

On the paramagnetic side, every local moment is fully screened.
The ground state is a global singlet.
Each local moment would contribute one Kondo resonance,
as inferred from the discussion of the previous subsection.
Since they have the quantum numbers of an electron, these Kondo 
resonances would combine with the conduction electron band.
The result is  a heavy quasiparticle band, with a 
``large'' Fermi surface: the volume that the Fermi
surface encloses counts the number of both the local moments
and conduction electrons.

The necessary condition for the Kondo resonances to form is
that the local moments are fully screened by the spins of
the conduction electrons, as we saw in the previous subsection.
This condition is equivalent to saying that the local spin
susceptibility must have a Pauli form. 
In other words,
the zero-temperature and zero-frequency limit of 
the local spin auto-correlation function must be finite.

Can this local spin susceptibility remain finite as the system 
approaches the magnetic quantum critical point? To address
this question, we note that in a translationally-invariant
system the local spin susceptibility is equal to the average
of the wavevector-dependent dynamical spin susceptibility
$\chi ({\bf q},\omega,T)$. Now, at an antiferromagnetic QCP,
by construction the peak susceptibility 
$\chi ({\bf Q},\omega,T)$ is divergent (where ${\bf Q}$ is the
antiferromagnetic ordering wavevector). When the peak susceptibility
diverges, the average susceptibility can either stay finite or 
become divergent as well.
If the average susceptibility is finite, then the local Kondo physics
proceeds without much modification.
On the other hand, a divergent average susceptibility
would be incompatible with the expected Pauli form of the local
spin susceptibility of a fully developed Kondo state.

This consideration already raises the possibility of two classes
of quantum critical metals.
For the more exotic type, the local spin susceptibility is divergent 
and fully developed Kondo resonances are absent at the quantum
critical point.
The heavy fermion physics then becomes a part of the critical theory.
This is a concrete example in which the emergence of non-Fermi liquid 
must be treated on an equal footing with the onset of magnetic ordering:
fermions cannot simply be taken as innocent bystanders.

\subsection{Local quantum critical points}

\begin{figure*}[h]
\centerline{\epsfysize=10cm\epsfbox{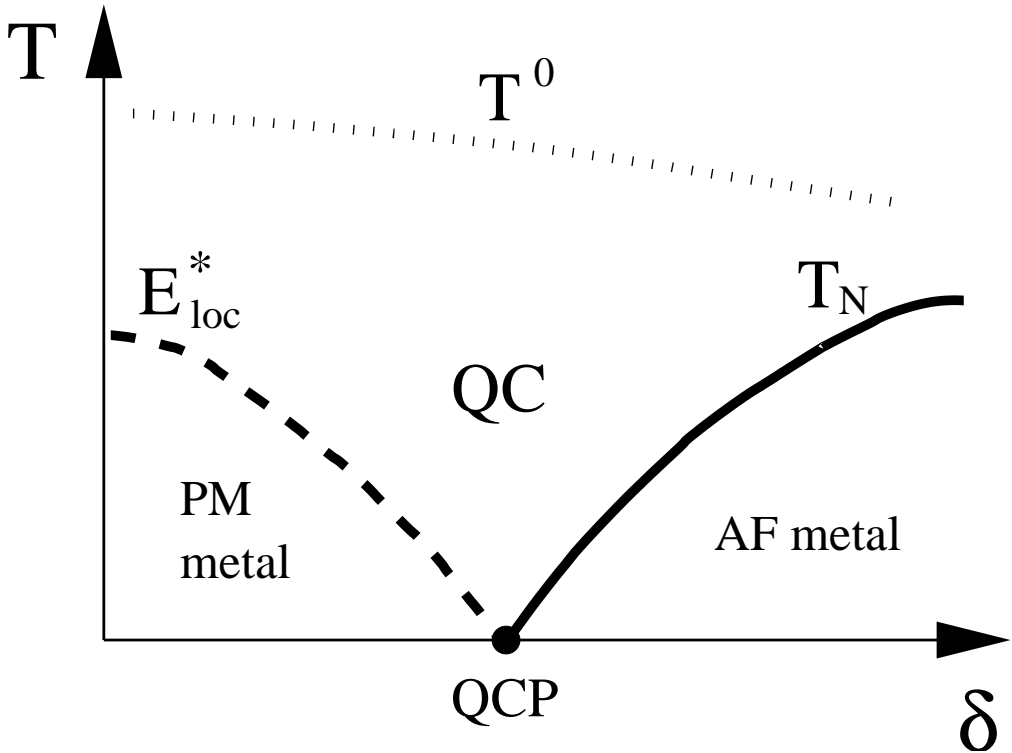}}
\caption{Local quantum critical point. Here $\delta$ is a tuning 
parameter.}
\label{lcp}
\end{figure*}

Since the Kondo effect and Kondo resonance formation are largely spatially
local phenomena, we have treated the interplay between the Kondo resonance
formation and onset of magnetic ordering within an 
extended dynamical mean field theory of the Kondo lattice model.
The more exotic type of quantum critical behavior arises 
when the magnetic fluctuations are two-dimensional. 
The schematic phase diagram is shown in Fig.~\ref{lcp}.
The vanishing of the 
energy scale $E_{loc}^*$ at the QCP signals the 
destruction of Kondo resonances
in the quantum critical regime:
the local susceptibility is Pauli only below $E_{loc}^*$.
If the magnetic fluctuations are
three dimensional, and if there is no magnetic 
frustration, then $E_{loc}^*$ 
would be finite at the QCP corresponding to
a crossover scale towards
the eventual Gaussian behavior.

In the quantum-critical regime, the local susceptibility is neither Pauli
nor Curie (the Curie behavior is recovered only above some 
cut-off scale $T^0$). It diverges logarithmically,
\begin{eqnarray}        
  \chi_{{loc}} (\omega) ~= ~{ 1 \over {2 \Lambda}}
         ~\ln {\Lambda \over {-i \omega}}
\label{chi-loc-lqcp}
\end{eqnarray}
where $\Lambda \approx T_K^0$. One can also introduce 
a ``spin self-energy'',
which captures the spin damping. It has the following form,
\begin{eqnarray}        
 M (\omega ) ~\approx &&-I_{\bf Q} + A ~(-i \omega)^{\alpha}
\label{M-lqcp}
\end{eqnarray}
Our recent numerical work for a Kondo lattice model
with an Ising anisotropy (done in collaboration with D. Grempel)
yields a nearly universal value for $\alpha$ that is 
close to $0.7$.

At finite temperature, the spin self-energy displays an $\omega/T$ scaling.
Both the fractional exponent and $\omega/T$ scaling reflect the non-Gaussian
nature of the fixed point. The corresponding critical theory captures both
the spatially-long-ranged fluctuations corresponding to the onset of 
magnetic ordering, and spatially local fluctuations reflecting the
destruction of Kondo resonances.
Based on Ginzburg-Landau considerations, we have also argued that
the results are robust
beyond the extended dynamical mean field theory.

\subsection{Experiments in heavy fermion metals}

The above analysis leads to a dynamical spin susceptibility
with $\omega/T$ scaling and, for every wavevector,
a fractional exponent ($\alpha<1$). 
The results are consistent with the inelastic neutron
scattering experiments of Schr\"oder and co-workers
mentioned earlier.

There is also a prediction that the NMR relaxation rate
contains a temperature-independent component. Recent
NMR experiments in ${\rm Yb Rh_2Si_2}$, by Kenji Ishida
and co-workers, have provided evidence for this.

Finally, the destruction of the Kondo resonance at the quantum
critical point implies 
that the Fermi surface undergoes a reconstruction at the QCP,
from being ``large''
(enclosing a volume that counts the local moments) to being 
``small'' (counting only the conduction electrons
and with a different topology) as the system orders.
There are de Haas-van Alphen experiments in heavy fermions
by Yoshichika \={O}nuki's group
which are suggestive of such a Fermi-surface reconstruction. 
Additional evidence is also
emerging from the Hall-coefficient measurements
of Silke Paschen and co-workers.

\section{Summary and outlook}

We have emphasized the important role electronic excitations
play in the quantum critical phenomena of correlated
metallic systems. Due to their non-Fermi liquid nature,
these electronic excitations may not simply serve as bystanders
as traditionally assumed in the Gaussian quantum critical metal
picture. Instead, they can become a part of the critical
degrees of freedom. The critical theory is then much richer than
the $\phi^4$ theory of paramagnons, opening the door to
non-Gaussian quantum critical metals.

Concrete progresses have been made in the context of 
quantum critical heavy fermions. Here, the electronic 
excitations are in fact the Kondo resonances.
In the local quantum critical
picture, these Kondo resonances are part of the critical
theory along with the paramagnons: a destruction
of the Kondo resonances accompanies an onset of magnetic
ordering. The non-Gaussian nature of the critical theory
allows fractional exponents and $\omega/T$ scaling. 
This picture is largely consistent with existing and
emerging experiments.

Some aspects of the heavy fermion phenomenology, 
such as $\omega/T$ scaling, have also been reported 
in high temperature cuprate superconductors,
including in the angle resolved photoemission
spectra over a wide range of wavevectors.
Microscopically,
a common feature between the cuprates and heavy fermions is
the strong Coulomb interaction, as manifested through
the formation of local magnetic moments in heavy fermions and
by the emergence of Mott insulating phase in the cuprates.
While quantum critical points in the cuprates have yet to be
explicitly identified, it appears hard to understand the 
scaling aspects of the phenomenology without invoking quantum
critical physics. In any event, to the extent that quantum 
criticality plays a role in the cuprates, the photoemission
results would suggest that the non-Fermi liquid electronic
spectrum are also a part of the critical theory.
As in the locally critical quantum phase transitions,
this can in turn make the quantum critical point non-Gaussian.


I would like to gratefully acknowledge D.\ Grempel, K.\ Ingersent, 
E.\ Pivovarov, S.\ Rabello, J.\ L.\ Smith, J.-X. Zhu and L.\ Zhu 
for collaborations,
APCTP and Professor H.\ Y.\ Choi for the invitation to write this
article, and NSF, TcSAM and Welch foundation for support.

\end{document}